\documentclass[aps,prb,twocolumn,showpacs]{revtex4}
\usepackage{graphics,amsmath}
\usepackage{tabularx,graphicx}
\usepackage{epsfig}

\begin{document}
\title{Phase diagram of the dissipative quantum particle in a box }
\author{J. Sabio$^1$}
\author{L. Borda$^{2,3}$}
\author{F. Guinea$^1$}
\author{F. Sols.$^4$}
\affiliation{$^1$Instituto de Ciencia de Materiales de Madrid (CSIC),
Cantoblanco, 28049 Madrid. Spain}
\affiliation{$^2$Research Group ``Theory of Condensed
    Matter'' of the Hungarian Academy of Sciences, TU Budapest, Budapest,
    H-1521, Hungary}
\affiliation{$^3$ Physikalisches Institut, Universit\"at Bonn, Nussallee 12, 
  D-53115 Bonn, Germany}
\affiliation{$^4$Departamento de F{\'\i}sica de Materiales. Universidad
  Complutense de Madrid. 28040 Madrid. Spain.}

\date{\today}

\begin{abstract}
We analyze the phase diagram of a 
quantum particle confined to a finite chain, 
subject to a dissipative environment described by an Ohmic spectral function.
Analytical and numerical techniques are employed to explore both the perturbative and non-perturbative regime of the model. For small dissipation the coupling 
to the
environment leads to
a narrowing of the density distribution, and to a displacement towards the
center of the array of accessible sites. 
For 
large
values of the dissipation, we find a
phase transition to a doubly degenerate phase which reflects the formation of
an inhomogeneous effective potential within the array. 
\end{abstract}
\maketitle
 
\section{Introduction.}

The problem of a quantum particle interacting with an environment deserves special attention since it has implications in fundamental areas as quantum measurement theory, quantum dissipation and quantum computation, among others. A quantum particle
interacting with an environment 
consisting of a continuum of
degrees of freedom, the Caldeira-Leggett model,\cite{CL83} is actually the
simplest model which can be used to study the destruction of
quantum coherence and the emergence of classical behavior in the framework of quantum mechanics. 

It is generally believed that the Caldeira-Leggett model captures the
essential features of the behavior of more complicated open quantum
systems. Therefore, its study as a toy model can be justified even if its not connected to any particular experimental realization. However, some extensions
of this model are known to be relevant in real systems: for instance when
analyzing dephasing in a qubit \cite{MSS01} or in a dissipative Josephson
junction.\cite{L87} Another application recently pointed out is the study of the decoherence induced in
mesoscopic systems by external gates.\cite{GJS04,G05,CSG06} This can be seen as a consequence of Caldeira-Leggett model reproducing the long time dynamics of particles interacting
with Ohmic environments.\cite{G03}

Three variations of this model have been particularly well studied: (i) The dissipative two level
system,\cite{C82,BM82,Letal87,W99} which is probably the most analyzed model in the context of quantum computation, being the archetype of a qubit in the presence of a bath.\cite{Nielsen}  (ii) A particle moving in a periodic potential,\cite{S83, FZ85} in a magnetic field,\cite{DS97} and in both a periodic potential and a magnetic field.\cite{CF92, NNG05} The case of a dissipative particle in a
periodic potential is relevant to the study of defects in
Luttinger liquids,\cite{KF92} while the case with both a magnetic
field and a periodic potential applies to a junction between more
than two Luttinger liquids.\cite{COA03} Finally, (iii) the dissipative free
particle \cite{HA85,GSI88} of interest in the study of quantum Brownian
motion. In the first two cases the system undergoes a phase transition, for a critical value of the coupling to the bath, to a phase where the particle is localized. This kind of transition, which belongs to the general class of boundary quantum phase transitions,\cite{V06} has been studied in the literature with many
different approaches including
 path integral, renormalization group and variational Ansatz.

In the present work, we study the phase diagram of a particle 
confined to a finite tight-binding chain 
coupled to
an 
Ohmic dissipative environment through its coordinate variable.
This is the simplest  
intermediate instance
between two of the limiting cases 
described above, the dissipative two level system and a particle in an
infinite array.  It represents a quantum particle
interacting with an ohmic reservoir whose motion is restricted to a finite
region. As discussed in detail below, the inclusion of the hard wall boundary conditions introduces inhomogeneities in the density distribution of the particle, and it yields a non-trivial phase
diagram, with a quantum critical point which can be characterized in
detail by using numerical techniques. 

The paper is organized as follows:
First we describe the model, and briefly review the main results of related models obtained in the past. Then, we discuss the main techniques employed to analyze the ground state of the system. The calculated phase diagram is discussed, as well as the way various
observable quantities are affected by the dissipation. The last part of the
paper summarizes the main results of the work.

\section{The model.}

As we have noted above, the model we address
is that of a particle coupled to a dissipative bath which can hop between $M$
sites. The Hamiltonian reads:
\begin{eqnarray}
  {\cal H} &\equiv &{\cal H}_{\mathrm{kin}} + {\cal H}_{\mathrm{bath}} + {\cal H}_{\mathrm{int}} +
  {\cal H}_{\mathrm{ct}} \nonumber \\
{\cal H}_{kin} &\equiv &-t \sum_{m=1}^{M} c_m^\dag c_{m+1} +
  \mathrm{h. c.} \nonumber \\
{\cal H}_{bath} &\equiv &\sum_{k < \omega_c}k b_k^\dag b_k \nonumber \\
{\cal H}_{int} &\equiv &\lambda 
q \sum_{k < \omega_c}
  \sqrt{k} \left( b_k^\dag + b_k \right) \nonumber \\
{\cal H}_{ct} &\equiv &\lambda^2 q^2 \sum_{k
  < \omega_c} 1 
\label{hamil}
\end{eqnarray}
Here $t$ is the hopping between nearest neighbor sites, and $\lambda$ the coupling with which the bath couples to the position of the particle $q \equiv  \sum_m (m-m_0)
  c_m^\dag c_m $, where $m_0$ labels the center of the chain. The bath is characterized
by a high energy cutoff $\omega_c$. The last term of the Hamiltonian is a counter-term introduced in order to preserve the degeneracy
between the energies of the different sites. We assume that the coupling leads to
Ohmic dissipation, so that:
\begin{equation}
J ( \omega ) = \pi \lambda^2 \sum_{k < \omega_c} k \delta ( \omega - k ) = 2 \pi
\alpha | \omega |\;, \qquad {\rm for} \quad \omega < \omega_c
\label{alpha}
\end{equation}
where $\alpha \equiv \lambda^2 / ( 4 \pi )$ and $J(\omega)$ is the spectral weight function.

This Hamiltonian can be transformed through a unitary operation into one in which all sites are treated on  the same footing because the transformed bath couples to the inter-site hopping. In order to arrive at this form, which can be useful in the study of the phase diagram, we use the transformation $U = e^{-\lambda q \sum_k \frac{1}{\sqrt{k}} (b^{\dagger}_k - b_k)}$ on the Hamiltonian, yielding:\cite{S83,HMG84}
\begin{equation}
{\cal H} = \sum_{k < \omega_c}k b_k^\dag b_k  - t \sum_{m=1}^M (c_m^{\dag} c_{m+1} e^{-\lambda \sum_k \frac{1}{\sqrt{k}} (b^\dag_k - b_k)} + h.c.)
\label{hamil2}
\end{equation}
Hamiltonians (\ref{hamil}) and (\ref{hamil2}) contain
two limiting cases of interest. The case $\lambda = 0$ is that of a confined
particle 
decoupled from the bath. The Hamiltonian is readily diagonalized, and the ground state corresponds to a particle delocalized with a density $\rho_m = \frac{2}{M+1}\sin^2(\frac{\pi m}{M+1})$. On the other hand, the case $t = 0$ corresponds to a particle without kinetic energy. From Hamiltonian (\ref{hamil2}) we see that the ground state is M-fold degenerate in the subspace of site states adiabatically dressed by a cloud of bosons, $|m\rangle\otimes e^{-m \lambda \sum_k \frac{1}{\sqrt{k}}(b^{\dagger}_k - b_k)}|0 \rangle$. In general, we will be mainly interested in generic values of the parameters $t$ and $\lambda$ of the Hamiltonian. Here, the bath can be regarded as performing repeated measurements of the position of the particle, localizing it in the sites basis, as opposed to the kinetic term, which tends to delocalize it. As in the dissipative two level system and the dissipative particle in a periodic potential, the phase diagram is expected to reflect the two opposing tendencies through a quantum phase transition.

In this work we will analyze the regime $t \ll \omega_c$, where we expect that the low-energy properties will depend only on the dimensionless parameters $t / \omega_c$ and $\alpha$. Notice that in the continuum limit, $M>>1$, where the couplings satisfy $\alpha <<1$,  the model should reduce to a particle described by an effective mass in a dissipative
  environment, which admits a complete analytical solution.\cite{HA85} For larger couplings we expect, as mentioned, a phase transition as seen in related models. To analyze this region we will concentrate in a small number of sites, ranging between 2 and 6, where both numerical and analytical calculations are easier to perform.

\section{The calculations.}

\subsection{Preliminary remarks.}

The existence of a phase transition for a large enough strength coupling can be seen simply by normal ordering Hamiltonian (\ref{hamil2}). This operation, which can be regarded as a resummation of an infinite series of tadpole diagrams for the interacting vertex,\cite{HMG84} gives a renormalized hopping $t_{ren} = t e^{-\frac{\lambda^2}{2} \sum_k \frac{1}{k}}$. Performing the summation we can write from this expression a flow equation for the hopping parameter in terms of the cutoff of the bath, giving the celebrated result:\cite{AGG80}
\begin{equation}
\frac{d}{d  \log \omega_c} (\frac{\Delta(\omega_c)}{\omega_c}) = (\alpha - 1)\frac{\Delta(\omega_c)}{\omega_c}
\end{equation}  
From this equation it follows that a phase transition exists for $\alpha = 1$
from a regime in which $t_{ren}$ is finite to one in which is effectively
renormalized to zero, thus suppressing quantum fluctuations. In general,
however, this is not the whole story and more precise calculations are needed
in order to get the right critical line, which can have a dependence on the
hopping parameter. This is the case, as is well known, for the dissipative two level system, where the transition can be shown to be of the Kosterlitz - Thouless (KT) kind \cite{K76} when higher corrections to the flow equations are computed.\cite{C82, BM82, HMG84}  On the contrary, for the dissipative particle in a periodic potential there are no higher order corrections to the flow equations and the transition line in the $\alpha - t$ phase diagram can be shown to be vertical.\cite{GHM85} 

A similar analysis can be tried for the dissipative confined particle, but the computation of higher order corrections to the transition results very tedious, because of the large number of coupled flow equations that must be analyzed. This is due to the lack of symmetries of this model (only parity is preserved), which generates an important number of counter-terms that must be taken into account in the critical region close to $\alpha = 1$. In particular we can have higher charges of the form
\begin{equation}
t_l \sum_{m=1}^{M}(c^{\dag}_m c_{m+l} e^{-\lambda l \sum_k \frac{1}{\sqrt{k}}(b_k^{\dag} - b_k)} + h.c.)
\nonumber
\end{equation}
where $l >1$, generating next-to-nearest neighbor hoppings and beyond in the low energy theory. However, by normal ordering this term can be shown to be irrelevant close to the transition. Also, a renormalization of the potential can be expected, $\sum_{m=i}^{M} v_m c_m^{\dag} c_m$, with $v_m = v_{-m}$ in order to preserve parity symmetry. And finally, a renormalization of $\lambda$ which we do not show here. 
The complexity of the problem justifies the application of numerical techniques such as the numerical renormalization group, which deal with the whole effective low-energy Hamiltonian.

\subsection{Variational Ansatz.}

Some deeper insight into the physics of the problem can be obtained by using a generalization of the variational Ansatz proposed by Silbey and Harris: \cite{SH89}
\begin{equation}
|G\rangle = e^{q \sum_k \frac{f_k}{k}(b_k^{\dag}-b_k)}|0\rangle\otimes \sum_m c_m |m\rangle
\label{ansatz}
\end{equation}
For the application to our problem we have enlarged the original set of variational parameters, $f_k$, in order to include the on-site amplitudes of the wave function, $c_m$. Without loss of generality, the latter are taken real. The energy of the above proposed ground state is:
\begin{equation}
E_G = 2 t_{ren} \sum_m c_{m+1} c_m + \langle q^2\rangle_G g
\end{equation}
where we have defined $t_{ren} \equiv t e^{-\frac{1}{2}\sum_k \frac{f_k^2}{k^2}}$ and $ g \equiv  \sum_k (\lambda + \frac{f_k}{\sqrt{k}})^2$.  The minimum condition imposes the following set of equations:
\begin{eqnarray}
t_{ren} (c_{m+1} + c_{m-1}) + (g (m-m_0)^2 - E_G) c_m = 0 \label{var1} \\
t_{ren}  \frac{f_k}{k^2} \sum_m c_{m+1} c_m - \langle q^2 \rangle_G (\frac{\lambda}{\sqrt{k}} + \frac{f_k}{k}) = 0
\end{eqnarray}
The first set of equations can be seen as those of a particle in a chain with
renormalized hopping $t_{ren}$ and moving in a symmetric 
parabolic
potential $v_m = g (m - m_0)^2$. The second equation is a self-consistent condition for the parameters $f_k$, once we have determined the lowest $E_G$ and the corresponding set of $c_m$ from (\ref{var1}).

\begin{figure}[tbh]
\begin{center}
\includegraphics*[ width=0.9\columnwidth]{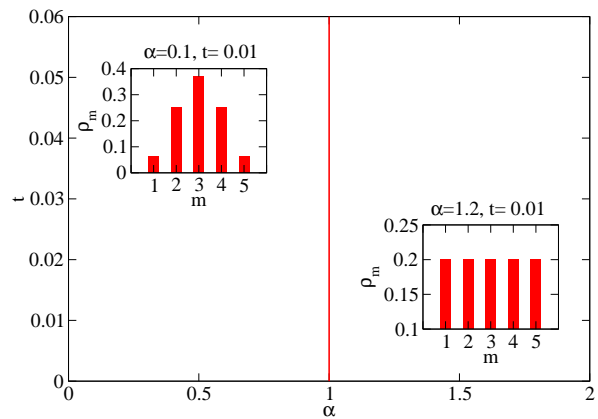}
\end{center}
\caption{Phase diagram predicted by the variational calculation, for a chain of 5 sites. The model shows a phase transition at the critical coupling $\alpha_c = 1$, where the parity symmetry is broken. For $\alpha < 1$ there is a delocalized phase, with $t_{ren}$ finite and an effective quadratic potential dependent on the coupling strength, that is responsible for the increasing localization of the particle at the central sites. For $\alpha > 1$ there is a localized phase, with $t_{ren} = 0$, and a M-fold degenerate ground state is predicted. Notice that, in the variational approach, a similar phase diagram is predicted for a chain of 6 sites.}
\label{var}
\end{figure}

\subsubsection*{Results}

The solution of these equations gives a phase transition for the critical
coupling $\alpha_c = 1$, as expected from the preceding
section. For $\alpha < 1$ the particle is not localized, as $t_{ren}$ is
nonzero, but a certain degree of localization at the center arises from the
existence of the potential term $g$. In the limit $\frac{t_{ren}}{\langle q^2 \rangle_G} \ll
\omega_c$, this factor can be shown to be $g = - 2
\frac{t_{ren}}{\langle q^2 \rangle_G} \alpha \sum_m c_{m+1} c_m$. As the coupling strength approaches the critical value, the localization of the particle in the central site becomes sharper, as can be seen in Fig. \ref{qcuad} for a chain of 5 sites. In this figure the position mean square of the particle is plotted as a function of the coupling strength. At the critical point the particle is completely localized in the center, as a consequence of the renormalization to zero of the hopping parameter and the appearance of the confining potential. 

For $\alpha > 1$ the particle gets localized, with $t_{ren} = 0$, and an $M$-fold 
degenerate ground state, in which the parity symmetry is broken, is predicted. This is, actually, the result of solving exactly the Hamiltonian (\ref{hamil2}) in the limiting case $t=0$, as discussed in Section I. The transition line predicted by this approach is vertical, not having any dependence on the hopping parameter. A phase diagram containing this features is shown in Fig. \ref{var}. 

The variational calculation shows that despite the Caldeira - Leggett model includes a counter-term to ensure homogeneous dissipation, the boundary conditions are responsible of some non homogeneous effects such as the creation of an effective parabolic potential which confines the particle at the center. However, as we will see in the next section, the numerical results suggest that the situation close to the critical point is more complicated, with new inhomogeneous terms playing an important role in the solution.

\subsection{Numerical Renormalization Group.}

In order to include higher correlation effects it is necessary to go beyond the variational solution. A powerful method to study quantum impurity problems is Wilson's numerical renormalization group (NRG).\cite{W75,BCP08} Originally conceived to deal with fermionic environments, it has been recently adapted to handle bosonic baths. \cite{BTV03,BLTV05} Bosonic environments can also be analyzed using the
well known correspondence between bosons and fermions in one
dimension,\cite{CK96,CM03,NNBZA03,NNBAZ05} 
used e.g. in the study of dissipative
gates.\cite{BZS05,BZG06} 
However, the fermionic model covers only a part of the
bosonic one, as the dissipation strength is limited to the range $0 < \alpha < 1$. Exactly
because of this fact, we use the bosonic version of the problem. 

\begin{figure}[tbh]
\includegraphics[  width=0.90\columnwidth]{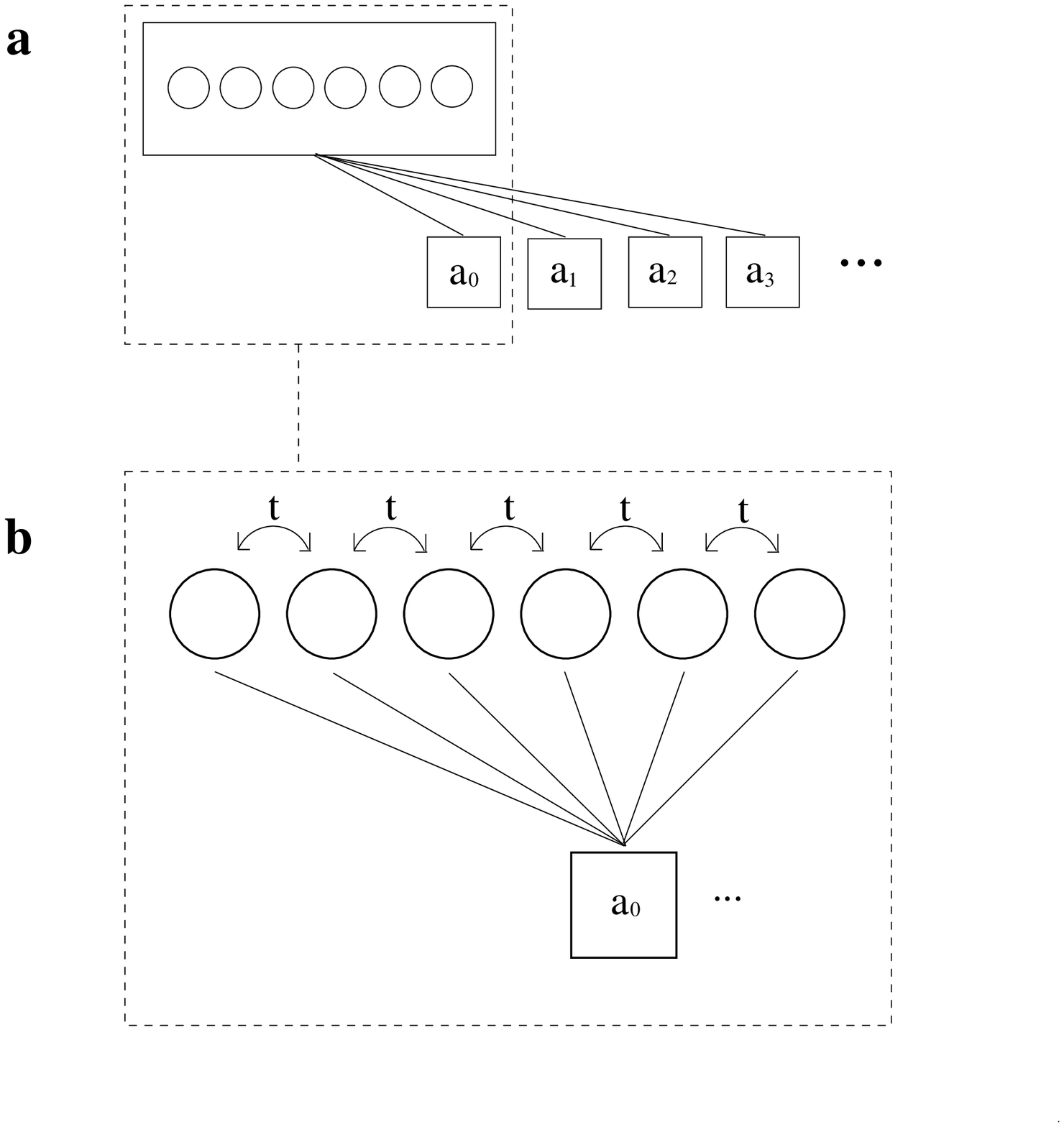}
\caption{Sketch of the star-NRG Hamiltonian used in the work. a) Every bosonic
  site $a_n$ couples to the particle. In each iteration a new site is added
  and the resulting Hamiltonian is diagonalized, giving the energy
  spectrum. b) The particle Hamiltonian, coupled to the first bosonic
  site. Notice that the structure of the couplings is the same for the rest of
  bosons. See the text for more details on the calculations.}
\label{method}
\end{figure}

When working with bosonic degrees of freedom, 
we use the so called star-NRG\cite{BLTV05}
method, schematically shown in Fig. \ref{method}. 
As opposed to the conventional ``chain'' NRG, this version allows us to deal with the counter-term in the Hamiltonian.
This issue arises due to the fact that the counter-term, which only involve particle operators, is of order of the cutoff. In the chain NRG, the Hamiltonian is transformed in such a way that all the particle dependence is included in the first iteration of the algorithm. The inclusion of the counter-term here requires much greater precision in the calculations, as opposed to the star-NRG, where is included iteratively. This is no longer the case in absence of the counter-term, as shown in Ref.\onlinecite{TTB06}. There the authors used the chain NRG to study dissipative exciton transfer.

The star-NRG method is based on the introduction of a new basis of bosonic states for which the couplings to the particle decrease exponentially as $\Lambda^{-n}$, where $\Lambda$ is a scaling factor between 2 and 3, and $n$ labels the bosonic sites. Then, the Hamiltonian is diagonalized iteratively, adding a new bosonic site in every step. 

The basis is truncated with both an upper cutoff in energies,
which is progressively reduced in each iteration with the scaling factor
$\Lambda$, and an upper cutoff in the occupation number of the bosonic sites. 
The first one is chosen to have a number of kept states in each iteration of
$N_s = 100 - 120$, while for the second we let $N_b = 30 - 40$ bosons per
site. We
check that the procedure gives well converged results, regardless of the chosen truncation parameters.
\begin{figure}[tbh]
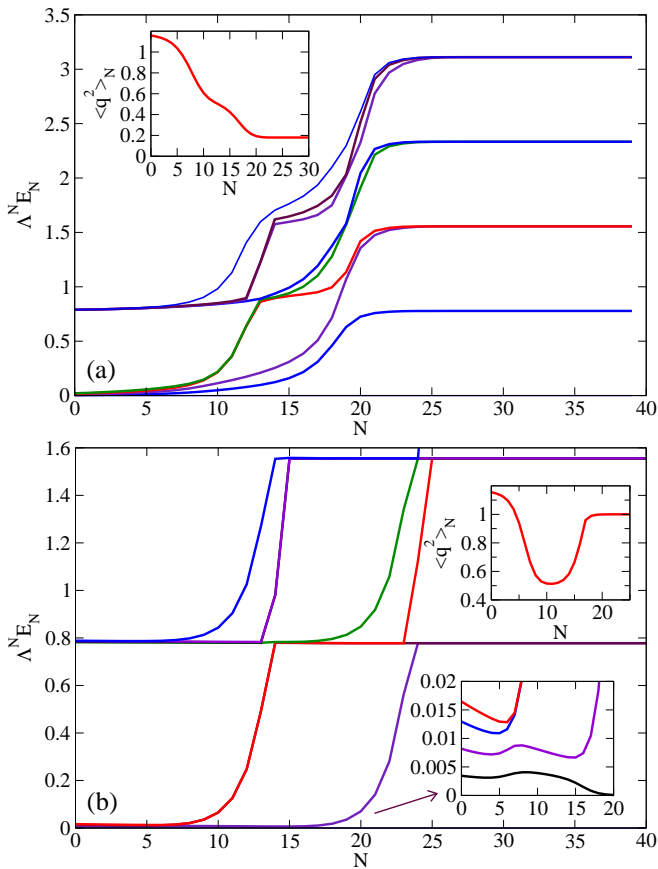

\begin{center}
\includegraphics*[  width=1\columnwidth]{3a.eps}
\includegraphics*[  width=1\columnwidth]{3b.eps}
\end{center}
\caption{Representative flows of the Numerical Renormalization Group
  transformations carried out in this work, for $M = 5$ sites. The horizontal axis is the
  iteration number, and the graphs are scaled energy levels. a) Flow towards a non
  degenerate ground state ($\alpha = 0.8$, $t = 0.01$) The inset shows the flow of the
  mean squared position of the particle, which becomes localized around the center. b) Flow towards a degenerate
  ground state ($\alpha = 1.2$, $t = 0.01$). The lower inset gives details of the way in
  which the lowest energy levels flow to the fixed point. The top inset shows the localization of the particle beyond the center, as indicated by its mean squared position. Notice that, in this regime, two energy scales are playing a part in the flow, delimitating an intermediate regime which is dominated by an unstable fixed point.}
\label{flow}
\end{figure}

\begin{figure}[tbh]
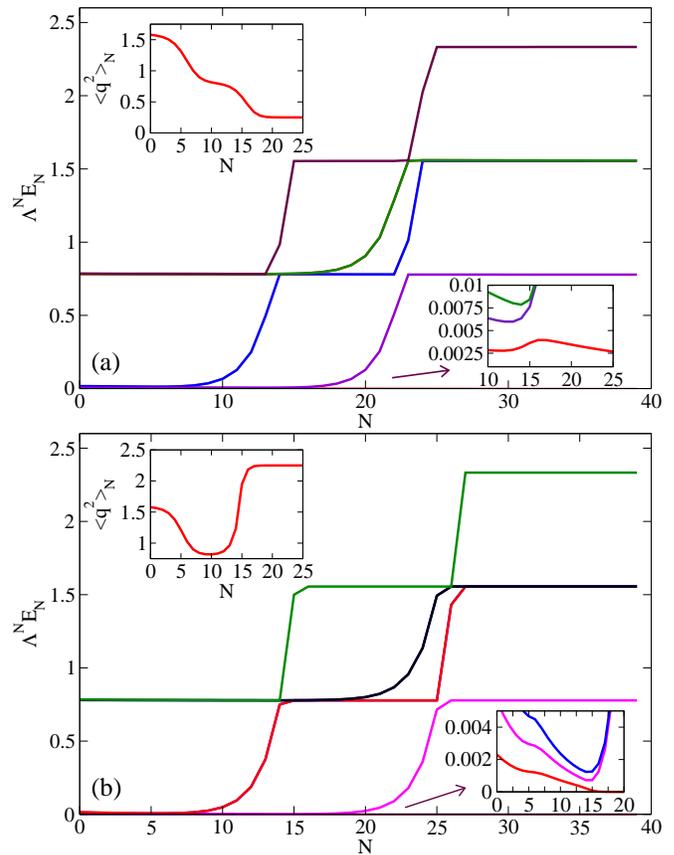

\begin{center}
\includegraphics*[  width=1\columnwidth]{4a.eps}
\includegraphics*[  width=1\columnwidth]{4b.eps}
\end{center}
\caption{Flows of the NRG for $M = 6$ sites. a) Flows in the first region of the localized phase ($\alpha = 1.2$, $t= 0.01$), where the particle is confined at the central sites. b) Flow in the second region of the localized phase  ($\alpha = 1.4$, $t= 0.01$). Here the particle is confined in the next to the center sites, suggesting the formation of a double well effective potential in the array. In both cases the lower insets show in detail the flow of the lowest energy levels. Notice that, as happened in the chain of five sites, two energy scales apparently delimit the onset of an unstable fixed point, corresponding to an effective chain of $M=4$ sites. This is actually well conformed by the value of the mean squared position of the particle (inset)}
\label{flow2}
\end{figure}

We exploit the parity symmetry of the Hamiltonian in the code, both for reducing the size of the matrices to be diagonalized and for minimizing numerical errors. Let $\Pi$ be the parity operator, under which the operators of the Hamiltonian transform as:
\begin{eqnarray}
\Pi c_m^{\dag} \Pi^{\dag} = c_{-m} \nonumber \\
\Pi b_k^{\dag} \Pi^{\dag} = -b_k^{\dag} \nonumber \\
\Pi q \Pi^{\dag} = -q  
\end{eqnarray}
Bosonic states have a well defined parity, $\Pi (b_k^{\dag})^{n_b} |0\rangle = (-1)^{n_b} (b_k^{\dag})^{n_b} |0\rangle$. The particle states must be rotated to the basis of eigenstates of parity, $c_{m,p} = \frac{1}{\sqrt{2}}(c_m + p c_{-m})$. For $M$ odd the central site is always guaranteed to be a parity eigenstate. The states used to diagonalize the NRG-Hamiltonian at zero iteration are then $|m,p;n_b;P\rangle$, with total parity $P = p(-1)^{n_b}$ being a good quantum number. The total Hamiltonian then splits into two separated sectors of well defined total parity, reducing by two the size of the matrices to be diagonalized. The same can be shown to be true at iteration $N+1$, where the total parity states $|r,p;n_b;P\rangle_{N+1}$ are constructed adding a new bosonic site to the eigenstates at iteration $N$ with parity $p$. The matrix elements of the NRG-Hamiltonian verify, in this basis:
\begin{equation}
\langle r',p';n_b';P'|{\cal H}_{N+1}|r,p;n_b;P\rangle \propto \delta_{P,P'} \nonumber
\end{equation}  
The output of the NRG procedure are the flows of the lowest lying energy states as the cutoff is reduced iteratively. At some point the flows are expected to converge to stable (low energy) fixed points. The effective Hamiltonian can be reconstructed analyzing the evolution of those flows, as well as the evolution of other observables of the system. Here we will use the evolution of the averaged position of the particle, $\langle q \rangle_N$, and its mean squared deviation $\langle q^2\rangle_N$, evaluated in the ground state. Those flows are enough to characterize the different phases of the system.

\subsubsection*{Results}

In this paper we analyze chains with a number of sites ranging between 2 and 6. For two sites the NRG reproduces the phase diagram of the dissipative two level system, as was shown by Bulla et al \cite{BLTV05}. For larger chains and small dissipation, the results are in quantitative agreement with the variational solution (see below), predicting a delocalized phase with renormalized hopping and a renormalized potential which tends to localized the particle at the center as the coupling strength is enlarged. As far as the phase transition is concerned, the case of $M=3,4$ does not deviate too much from the dissipative two level system. In both cases there is a phase transition in which $t_{ren} = 0$, but for $M=3$ the parity symmetry is not broken, because the particle is localized at the center. For $M=4$ the phase transition is that of the dissipative two level system, the edge sites being decoupled in energy from the central ones. This is all in contrast with the variational solution, where a transition to an $M$ degenerate state is predicted.

\begin{figure}[tbh]
\begin{center}
\includegraphics*[  width=0.9\columnwidth]{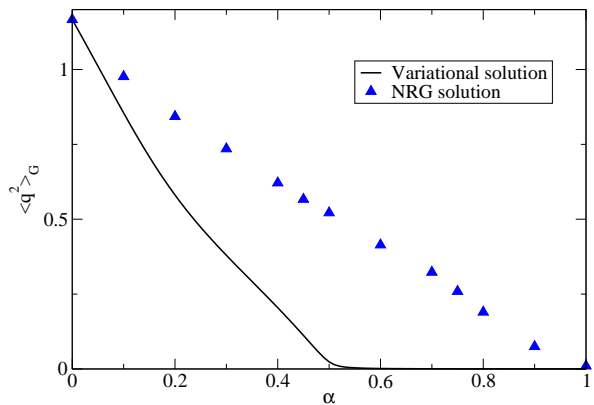}

\end{center}
\caption{Mean-squared position of the particle as a function of the coupling strength in the delocalized phase of the model, for M = 5, as predicted by the variational calculation and the NRG. Both approaches show an increasing localization of the particle at the center of the chain as the coupling strength gets larger. This effect arises due to the renormalization of the hopping parameter and the emergence of an effective confining potential. In the variational calculation the localization is stronger, suggesting that higher order corrections to the effective potential are playing an important role in the numerical calculation.}
\label{q}
\label{qcuad}
\end{figure}

Of more interest are the cases of $M=5,6$. Here a new behavior is observed, which should be representative of the one expected for larger chains. The energy flows for small and large dissipation are shown in Fig. \ref{flow} and Fig. \ref{flow2}. Again, weak dissipation induces some localization at the center of the array of the particle density, as can be seen in the inset of the figures, where $\langle q^2 \rangle_N$ is computed. As mentioned above, in this regime the results agree qualitatively with the variational solution, as shown in Fig. \ref{qcuad}, where the mean squared position of the particle is calculated in both approximations for a chain of 5 sites (a similar plot can be obtained for 6 sites, the main difference being that the mean squared position tends to a finite value for $\alpha \rightarrow 1$, having two states in the center instead of one).

The differences between the localized phase predicted by the variational calculation and the NRG are even sharper for those longer chains. Above a critical strength coupling $\alpha_c$ of order one, we find a doubly degenerate state, for odd and even number of sites, in which the parity symmetry is broken. For $M=5$ the particle localizes next to the center, while for $M = 6$ it does initially in the central sites, and for larger dissipation in the next to the center ones. This result follows from analyzing the degeneracy of the ground state, extracted from the energy flows, as well as the evolution of the mean squared position operator. The converged values of the latter can be used to make an ansatz of the sort of ground state density matrix to which the flow converges. 

\begin{table}

\begin{tabular}{||l||c|c|c|c|c|}
\hline \hline  Type &M & $GS_{deg}$ & $\langle q^2 \rangle_{NRG}$ &$\hat{\rho}$ \\
 \hline
Deloc. &  5 & 1 &$\langle q^2 \rangle_{NRG}$ & $\hat{\rho} = |t_{ren}, \alpha \rangle \langle t_{ren}, \alpha|$   \\
Loc.   & 5 & 2 &1 & $\hat{\rho} = \frac{1}{2}(|2\rangle \langle 2| + |4\rangle \langle 4|)$ \\
Deloc.  & 6 & 1 & $\langle q^2 \rangle_{NRG}$ &  $\hat{\rho} = |t_{ren},\alpha\rangle \langle t_{ren},\alpha|$   \\
Loc. I & 6 & 2 &0.25& $\hat{\rho} = \frac{1}{2}(|3\rangle \langle 3| + |4\rangle \langle 4|)$  \\
Loc. II  & 6 & 2 &2.25& $\hat{\rho} = \frac{1}{2}(|2\rangle \langle 2| + |5\rangle \langle 5|)$  
\\ \hline \hline
\end{tabular}
\caption{Stable fixed points of the NRG for chains of $M = 5,6$ sites. The fixed points are characterized by the ground state degeneracy, $GS_{deg}$ and the position mean squared value, $\langle q^2 \rangle_{NRG}$, whose values can be obtained with the NRG (in the localized phase there is a single value for every $\alpha$, while in the delocalized one the value depends on the coupling strength, as shown in Fig. 5 for a chain of 5 sites).  Those are used to propose an ansatz for the ground state density matrix, $\hat{\rho}$, which fits it correctly. The states $|i\rangle$ represent a particle sitting at site $i$. $|t_{ren},\alpha\rangle$ is the ground state of a free tight-binding chain with hopping $t_{ren}$ and a parabolic potential, dependent on the strength coupling $\alpha$. The latter is the output of the variational calculation for the delocalized phase, which fits quite well the numerical data. } \label{table1}
\end{table}

In Table \ref{table1}, the zoo of stable fixed points of the model is shown, for chains of $M = 5,6$ sites. In the case of $M=6$ there is an extra fixed point in the localized regime, corresponding to a situation in which the particle finds more favorably to get localized in the sites next to the edges than next to the central sites. This second transition also occurs for a critical value of the coupling strength, but there is neither a symmetry breaking nor a change in the degeneracy of the ground state. Thus, the information provided by NRG is not enough to fully characterize the nature of this phase transition.

 This is not the only limitation of the numerical method. It also does not allow us to study large values of the
dissipation, $\alpha \gg 1$, as the occupancies of the bosonic states become
also high. The question of whether other phase transitions can be ruled out for high values of $\alpha$ remains open, and deserves a separate study with different techniques.

\begin{figure}[tbh]
\begin{center}
\includegraphics*[  width=0.9\columnwidth]{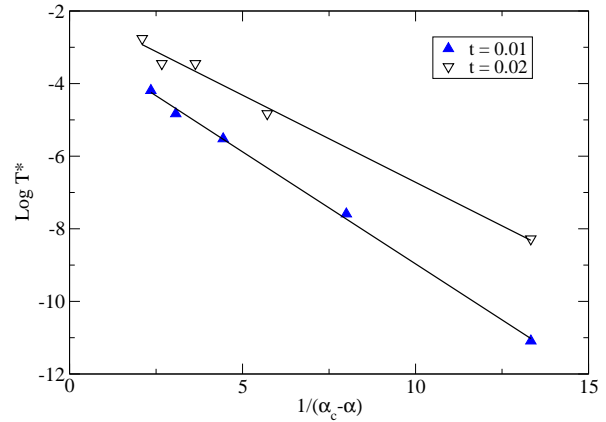}

\end{center}
\caption{Plot of the dependence of the crossover scale $T^*$ on the distance to the critical coupling strength. Here $T^* \propto \Lambda^{-N^*}$, with $N^*$ chosen as the iteration for which the first excited level verifies $\Lambda^{N^*}E_{N^*, 1} = 0.03$. The figure shows the results for two different hopping parameters. In both cases there is a good agreement with an exponential decay of $T^*$ as a function of the distance from the critical coupling, $\log T^* \propto 1/(\alpha_c - \alpha)$ }
\label{KT}
\end{figure}

From the energy flows some extra information can be extracted. In the delocalized phase, a single energy scale seems to be playing a part in the evolution from high energy to low energy behavior. Actually, the flow in this phase is similar to that in the dissipative two
level system, and in the same way we can define a crossover scale $T^* \propto \Lambda^{-N^*}$ from the iteration $N^*$ at which the flow
changes from its initial behavior to the low energy regime. As shown in Fig. \ref{KT}, $T^*$ tends to zero exponentially as the coupling strength approaches the critical value, $\log T^* \propto 1/(\alpha_c - \alpha)$. Hence, our results suggest that the transition is continuous, being consistent with the existence of a KT transition.\cite{BTV03}
 
The flows in the localized phase show a different behavior. Here, two different energy scales appear in the course of the flow, revealing an unstable fixed point in an intermediate regime. Those scales are defined now from the iterations $N_1^*$ and $N_2^*$ at which the energy levels decouple from the low energy sector, giving rise to two crossover temperatures $T_i \propto \Lambda^{-N_i^*}$, $i= 1,2$. From the mean squared position flows it can be deduced that the upper energy scale marks the decoupling of the sites located at the edges, as the values of this operator are well fitted to the expected ones in free tight-binding chains with finite hoppings but sites in the edges supressed (effectively reducing the chain to one with two less sites). In this way, the intermediate fixed point would correspond to an effective cluster of three sites in the $M=5$ case, and four sites in the $M = 6$ one, with a renormalized hopping parameter $t_{ren}$. The lower energy scale corresponds to the onset of the phase transition, for here the parity symmetry is broken and only two sites remain in the low energy regime.  

\section{Phase diagram}

\begin{figure}[tbh]
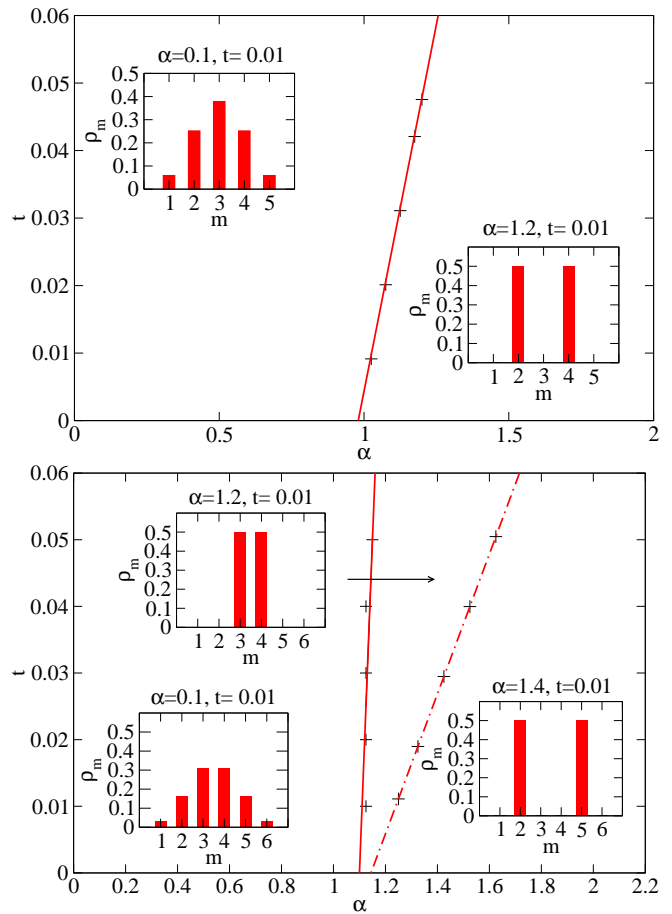

\includegraphics*[  width=1\columnwidth]{7a.eps}
\includegraphics*[  width=1\columnwidth]{7b.eps}
\caption{Phase diagram of the model, eq.(\protect{\ref{hamil}}), for $M=5$
  (top) and $M=6$ (bottom) sites, deduced from the NRG flow. The continuous lines are linear fittings to the numerical data. The insets show representative density distributions
  of the particle in each phase. In both cases there is a phase transition to a localized phase in which the parity symmetry is broken. The even case shows also a second transition in the localized phase, where the particle is confined beyond the central sites, resembling the kind of localization observed in the odd case.}
\label{phase_diagram}
\end{figure}

A phase diagram of the model,
including all the features discussed above, is presented in
Fig. \ref{phase_diagram}. As in the widely studied dissipative two level
system, there is a phase transition between a delocalized regime to a
localized one. In the delocalized phase the effect of the bath is that of reducing the
effective hopping and of generating a renormalized potential which makes the density of the particle higher close to the center. In the localized phase the parity symmetry is broken and in both cases, odd and even, the particle localizes in one of two degenerate sites.  This transition is continuous, and the numerical results are consistent with a
transition of the KT type, as in similar dissipative
systems.  
 
The variational calculation shows how a parabolic effective potential emerges from the coupling to the bath, being responsible of some localization of the particle at the center. However, the numerical results suggest the existence of a more complicated renormalized potential which would explain both the almost complete localization of the particle as $\alpha \rightarrow 1$ from below, and the inhomogeneous degenerate ground state in the localized phase. A simple guess which works well qualitatively is a effective potential in the form $V_m = (g_0/2)(m - m_0)^2 + (g_1/4!)(m - m_0)^4$. If $g_0 \propto (\alpha_c - \alpha)$, this Ansatz has a minimum at $m = m_0$ in the delocalized phase, and at $m = m_0 \pm \sqrt{\frac{-6 g_0}{g_1}}$ in the localized one, explaining not only the doubly degenerate state, but also that in the case $M = 6$ there is a second transition to a phase in which the particle localizes in the next to the center sites. From the point of view of renormalization theory, such an ansatz is reasonable as higher corrections to the potential should be more irrelevant.

In the region of the phase diagram which we have analyzed ($\alpha < 2$), we have not found any further crossover to a region where the particle is confined to the edges. This could be explained by the role played by the intermediate unstable fixed point in the localized phase. By studying the case of three and four sites, the only way to get a phase transition in which the particle localizes at the edges is by starting with slightly lower site energies here as compared to the center. Hence, the decoupling of the edges would be necessary to give rise to such a renormalization of the on-site energies, favoring the phase transition to a more stable regime. 

For larger chains we expect a similar picture to work, and new features should not appear as far as the continuous limit is not reached. The phase diagram should show a transition to a localized phase for a critical value $\alpha_c$ also around the unity. Close to this phase transition, the renormalized potential gets quartic corrections and the particle becomes localized in the resulting double-well profile. As the minimum of this effective potential depends on the dissipation strength, there should be several crossovers to regions in which the particle is localized at points increasingly farther from the center. However, the particle is never localized at the edges, since their decoupling seems to be crucial to the realization of this inhomogenous transition. Thus, in the localized phase, two or more energy scales are expected to play a role, depending on the number of energy levels that are decoupled from the low-energy sector until the stable fixed point is reached. In the continuous model, which corresponds to the case of an infinite number of sites in the array, we expect only a single phase transition, at $\alpha_c$ around the unity, to a phase where the particle is localized in a double well potential profile whose minimum depends on the coupling strength. 

\section{Conclusions.}
We have analyzed the simplest extension of the widely studied dissipative two
level system, which interpolates between this model and the also extensively
analyzed dissipative quantum particle in a periodic potential. The main results are obtained with 
a numerical renormalization group technique specifically adapted to bosonic
coordinates. For small and intermediate values of the dissipation parameter,
$\alpha$, our results provide a well controlled approximation to the ground
state of the system.

When dissipation is weak, we find that the density distribution of the
particle becomes narrower, and localized around the center of the array. This
result is consistent with the expectation that the environment acts as a
measurement apparatus on the position of the particle, leading to a more
localized distribution. The initial degeneracy between the sites of the array
is broken by the combination of the abrupt boundary conditions and the
dissipation. As a result, an effective local potential is induced, with a
minimum at the center of the array, leading to the confinement of the
particle near the center. 

For values of the dimensionless dissipation strength $\alpha \gtrsim 1$, we
find a transition to a situation with a doubly degenerate ground state, in which the parity symmetry is broken. In this case, the combined effect of boundary conditions and dissipation leads to the formation of an effective double well potential irrespective of the number of sites in the array. 

The formation of an  inhomogeneous effective potential from a spatially homogeneous coupling to a dissipative bath should be a generic feature in
similar models. A more remarkable result is that, in some cases,
this effective potential is non-monotonic, so that simple confinement
geometries can lead to complicated patterns in the localized regime for high
values of the dissipation. An open
question is to what extent these results may depend on the
particular choice of particle-bath coupling. It has been argued in
Ref. \onlinecite{ScS94} that the introduction of a counter-term is
insufficient to introduce full translational invariance in some
dynamic contexts, such as that of a suddenly introduced coupling. An
extension of the present equilibrium study to models of truly
translationally invariant dissipation \cite{ScS94} could shed some
light on this issue.

\section{Acknowledgements.}
We acknowledge A. J. Leggett for valuable discussions. This work was supported by MEC (Spain) through
grants FIS2005-05478-C02-01, FIS2004-05120 and FIS2007-65723, the Comunidad de Madrid,
through the program CITECNOMIK, CM2006-S-0505-ESP-0337, and EU Marie Curie RTN Programme no. MRTN-CT-2003-504574. J.S. wants to
acknowledge the I3P Program from the CSIC for funding. 
L.B. acknowledges the financial support 
of the J\'anos Bolyai Foundation, 
 the Alexander von Humboldt Foundation and 
Hungarian Grants OTKA
through projects T048782 and K73361.
\bibliography{text}

\end{document}